\documentclass[12pt]{article}
\usepackage[centertags]{amsmath}
\usepackage{amssymb,amsfonts}
\usepackage[comma,super]{natbib}
\usepackage{latexsym}
\newcommand{\be}{\begin{equation}} \newcommand{\ee}{\end{equation}}
\newcommand{\bea}{\begin{eqnarray}} \newcommand{\eea}{\end{eqnarray}}
\newcommand{\bse}{\begin{subequations}} \newcommand{\ese}{\end{subequations}}

\begin{document}

\begin{center}
{\Large \bf Charged relativistic spheres with generalized potentials} \\
\vspace{1.5cm} {\bf S. Thirukkanesh$^\dag$ and S. D. Maharaj}\\
Astrophysics and Cosmology Research Unit,\\
School of Mathematical Sciences,\\
University of KwaZulu-Natal,\\
Private Bag X54001,\\
Durban 4000,\\
South Africa.\\
\vspace{1.5cm} {\bf Abstract}\\
\end{center}
A new class of exact solutions of the Einstein-Maxwell system is
found in closed form. This is achieved by choosing a generalised
form for one of the gravitational potentials and a particular form
for the electric field intensity. For specific values of the
parameters it is possible to write the new series solutions in terms
of elementary functions. We regain well known physically reasonable
models. A physical analysis indicates that the model may be used to
describe a charged sphere. The influence of the electromagnetic
field on the
gravitational interaction is highlighted.
~\\
AMS classification nos: 83C15; 83C22; 85A99\\
~\\
$^\dag$Permanent address: Department of Mathematics, Eastern
University, Sri Lanka, Chenkalady, Sri Lanka.

\newpage
\section{Introduction}
Solutions of the Einstein-Maxwell system of equations for static
spherically symmetric interior spacetimes are necessary to
describe charged compact objects in relativistic astrophysics
where the gravitational field is strong as in the case of neutron
stars. In the presence of an electromagnetic field, the
gravitational collapse  of a spherically symmetric distribution of
matter to a point  singularity may be avoided: the
 gravitational attraction is counterbalanced by the Coulombian
 repulsive force in addition to the pressure gradient. The recent
 analyses of Ivanov [1] and Sharma {\em et al\/} [2]
show that the presence of the  electromagnetic field affects the
values of redships, luminosities and maximum mass of a compact
relativistic object. Patel and Koppar [3], Tikekar and Singh [4],
Mukherjee [5] and Gupta and Kumar [6] demonstrated that it is
possible to model charged neutron stars with high densities with
acceptable  bounds for the surface redshift, luminosity and total
mass. Thomas {\em et al\/} [7], Tikekar and Thomas [8] and Paul and
Tikekar [9] demonstrated that charged relativistic solutions may be
applied to core-envelope models. The role of the electromagnetic
field in describing the gravitational behaviour of stars composed of
quark matter has been recently highlighted by Mak and Harko [10] and
Komathiraj and Maharaj [11]. Therefore, the Einstein-Maxwell system
for a charged star has attracted considerable attention in various
physical investigations. We note that at present there is no unified
theory of electromagnetism and gravitation which is generally
accepted. In this paper we have used the approach of coupling the
electromagnetic field tensor to the matter tensor in Einstein's
equations such that Maxwell's equations are satisfied. We believe
that the qualitative features generated in this charged model should
yield results which are physically reasonable.

Our objective is to generate a new class of solutions to the
Einstein-Maxwell system that satisfies the physical criteria: the
gravitational potentials, electric field intensity and matter
variables must be finite and continuous throughout the stellar
interior. The speed of the sound must be  less than the speed of the
light, and ideally the solution should be stable with respect to
radial perturbations. A barotropic equation of state, linking the
isotropic pressure to the energy density, is often assumed to
constrain the matter distribution. In addition to these conditions
the interior solution must  match smoothly at the boundary of the
stellar object with the Reissner-Nordstrom exterior spacetime. In
recent years researchers have attempted to introduce a systematic
approach to finding solutions to the field equations. Maharaj and
Leach [12]  generalised the Tikekar superdense star, Thirukkanesh
and Maharaj [13] generalised the Durgapal and Bannerji neutron star
and Maharaj and Thirukkanesh [14] generalised the John and Maharaj
[15] model.  These new classes of models were obtained by reducing
the condition of pressure isotropy to a recurrence relation with
real and rational coefficients which could be solved by mathematical
induction, leading to new mathematical and physical insights in the
Einstein-Maxwell field equations. We attempt to perform a similar
analysis here to the coupled Einstein-Maxwell equations for a
general form of the gravitational potentials with charged matter. We
find that the generalised condition of pressure isotropy leads to a
new recurrence relation which can be solved in general.

In this paper, we seek new exact solutions to the Einstein-Maxwell
field equations, using the systematic series analysis, which may be
used to describe the interior relativistic sphere. Our objective is
to obtain a general class of exact solutions which contains
previously known models as particular cases. This approach produces
a number of difference equations, which we demonstrate can be solved
explicitly from first principles. We first express the
Einstein-Maxwell system of equations for static spherically
symmetric line element as an equivalent system using the Durgapal
and Bannerji [16] transformation in Section 2. In Section 3, we
choose particular forms for one of the gravitational potentials and
the electric field intensity, which reduce the condition of pressure
isotropy to a linear second order equation in the remaining
gravitational potential. We integrate this generalised condition of
isotropy equation using Frobenius method in Section 4. In general
the solution will be given in terms of special functions. However
elementary functions are regainable, and in Section 5, we find two
category of solutions in terms of elementary functions by placing
certain  restriction on the parameters. In Section 6, we regain
known charged Einstein-Maxwell models and uncharged Einstein models
from our general class of models. In Section 7, we discuss the
physical features, plot the matter variables and show that our
models are physically reasonable. Finally we summarise the results
found in this paper in Section 8.

\section{The Field Equations}
 The gravitational field should be static and
spherically symmetric for describing the internal structure of a
dense compact relativistic sphere which is charged. For describing
such a configuration, we utilise coordinates $(x^{a}) =
(t,r,\theta,\phi),$ such that the generic form of the line element
is given by
\begin{equation}
\label{eq:b1} ds^{2} = -e^{2\nu(r)} dt^{2} + e^{2\lambda(r)}
dr^{2} +
 r^{2}(d\theta^{2} + \sin^{2}{\theta} d\phi^{2}).
\end{equation}
The Einstein field equations can be written in the form
\begin{subequations}
\begin{eqnarray}
\label{eq:b2a}
\frac{1}{r^{2}} \left[ r(1-e^{-2\lambda}) \right]' & = & \rho \\
&  & \nonumber \\
- \frac{1}{r^{2}} \left( 1-e^{-2\lambda} \right) +
\frac{2\nu'}{r}e^{-2\lambda} & = & p \\
&  & \nonumber \\
\label{eq:b2c} e^{-2\lambda} \left( \nu'' + \nu'^{2} +
\frac{\nu'}{r} -\nu'\lambda' - \frac{\lambda'}{r} \right) & = & p
\end{eqnarray}
\end{subequations}
for neutral perfect fluids. The energy density $\rho$ and the
pressure $p$ are measured relative to the comoving fluid
4-velocity $u^a = e^{-\nu} \delta^a_0$ and primes denote
differentiation with respect to the radial coordinate $r$. In the
system (\ref{eq:b2a})-(\ref{eq:b2c}), we are using units where the
coupling constant $\frac{8\pi G}{c^4}=1$ and the speed of light
$c=1$. This system of equations determines the behaviour of the
gravitational field for a neutral perfect fluid source. A
different but equivalent form of the field equations can be found
if we introduce the transformation
\begin{equation}
\label{eq:b3} x = Cr^2,~~~~ Z(x)  = e^{-2\lambda(r)}, ~~~~
A^{2}y^{2}(x) = e^{2\nu(r)}
\end{equation}
so that the line element (\ref{eq:b1}) becomes
\[ ds^2 = -A^2 y^2 dt^2 + \frac{1}{4CxZ}dx^2 +
\frac{x}{C} (d\theta^2 +\sin^2\theta d\phi^2).\]
The parameters  $A$ and $C$ are arbitrary constants in
(\ref{eq:b3}). Under the transformation (\ref{eq:b3}), the system
(\ref{eq:b2a})-(\ref{eq:b2c}) has the equivalent form
\begin{subequations}
\begin{eqnarray}
\label{eq:b4a}
\frac{1-Z}{x} - 2\dot{Z} & = & \frac{\rho}{C} \\
&  & \nonumber \\
4Z \frac{\dot{y}}{y} + \frac{Z-1}{x} & = & \frac{p}{C} \\
&    & \nonumber\\
\label{eq:b4c} 4Zx^{2}\ddot{y} + 2 \dot{Z}x^{2} \dot{y} +
(\dot{Z}x - Z + 1)y & = & 0
\end{eqnarray}
\end{subequations}
where dots represent differentiation with respect to $x$.

In the presence of an electromagnetic field the system
(\ref{eq:b4a})-(\ref{eq:b4c}) has to be replaced by the
Einstein-Maxwell system of equations. We generate the system
\begin{subequations}
\begin{eqnarray}
\label{eq:b5a} \frac{1-Z}{x} - 2\dot{Z} & = & \frac{\rho}{C} +
 \frac{E^{2}}{2C} \\
& & \nonumber \\
4Z\frac{\dot{y}}{y} + \frac{Z-1}{x} & = & \frac{p}{C} -
 \frac{E^{2}}{2C} \\
&  & \nonumber \\
\label{eq:b5c}4Zx^{2}\ddot{y} + 2 \dot{Z}x^{2} \dot{y} +
\left(\dot{Z}x -
Z + 1 - \frac{E^{2}x}{C}\right)y & = & 0 \\
&  & \nonumber \\
\label{eq:b5d} \frac{\sigma^{2}}{C} & = & \frac{4Z}{x} \left(x
\dot{E} + E \right)^{2}
\end{eqnarray}
\end{subequations}
where $E$ is the electric field intensity and $\sigma$ is the
charge density. This system of equations governs the behaviour of
the gravitational field for a charged perfect fluid source. When
$E=0$ the Einstein-Maxwell equations (\ref{eq:b5a})-(\ref{eq:b5d})
reduce to the uncharged Einstein equations
(\ref{eq:b4a})-(\ref{eq:b4c}).

\section{Choosing $Z$ and $E$}
We  seek solutions to the
Einstein-Maxwell
 field equations (\ref{eq:b5a})-(\ref{eq:b5d}) by making explicit
 choices for the gravitational
potential $Z$ and the electric field intensity $E$ on physical
grounds. The system (\ref{eq:b5a})-(\ref{eq:b5d})
 comprises four equations in six unknowns $Z, y, \rho, p, E$ and $\sigma$.
 Equation (\ref{eq:b5c}), called the generalised condition of pressure isotropy, is the
 master equation in  the integration process.
In this treatment we specify the gravitational potential $Z$ and
electric field intensity $E$, so that it is possible to integrate
(\ref{eq:b5c}).  The explicit solution of the Einstein-Maxwell
system (\ref{eq:b5a})-(\ref{eq:b5d}) then follows.  We make the
particular choice
\begin{equation}
\label{eq:b6} Z = \frac{1+a x}{1+b x}
\end{equation}
where $a$ and $b$ are  real constants. The function $Z$ is regular
at the centre and well behaved in the stellar interior for a wide
range of values of $a$ and $b$. It is important to note that the
choice (\ref{eq:b6}) for $Z$ is is physically reasonable. This form
for the potential $Z$ contains special cases which correspond to
neutron star models, eg. when $a= - \frac{1}{2}$ and $b=1$ we regain
the uncharged dense neutron star of Durgapal and Bannerji [16]. When
$a$ is arbitrary and $b=1$ then Thirukkanesh and Maharaj [13] and
Maharaj and Komathiraj [17] found charged solutions to the
Einstein-Maxwell system. These solutions can be used to model a
charged relativistic sphere with desirable physical properties.
Consequently the general form (\ref{eq:b6}) contains known
physically acceptable uncharged and charged relativistic stars for
particular values of $a$ and $b$. We seek to study the
Einstein-Maxwell system with the choice (\ref{eq:b6}) in an attempt
to find new solutions, and to show explicitly that cases found
previously can be placed into our general class of models.

Upon substituting (\ref{eq:b6}) in equation (\ref{eq:b5c}) we
obtain
\begin{equation}
\label{eq:b7} 4(1+a x)(1+bx) \ddot{y}+2(a-b) \dot{y}+
\left[b(b-a)-\frac{E^{2}(1+bx)^{2}}{Cx}\right]y = 0.
\end{equation}
The differential equation (\ref{eq:b7}) is difficult to solve in
the above form; we first introduce a transformation
 to obtain  a more convenient form. We let
\begin{equation}
\label{eq:b8} 1+b x  = X,~~ Y(X) = y(x), ~~b \neq 0.
\end{equation}
With the help of  (\ref{eq:b8}),  (\ref{eq:b7}) becomes
\begin{equation}
\label{eq:b9} 4X \left[a X-(a-b)\right]\frac{d^{2}Y}{dX^{2}} +
2(a-b)\frac{dY}{dX} +
 \left[ (b-a)-  \frac{E^{2}X^{2}}{C(X-1)} \right]Y = 0
\end{equation}
in terms of the new dependent and independent variables $Y$ and
$X$ respectively. The differential equation (\ref{eq:b9})
 may be integrated once  the electric field $E$ is given.
A variety of choices for $E$ is possible but only a few have the
desirable features in the stellar interior. Note that the
particular choice
\begin{equation}
\label{eq:b10} E^{2} = \frac{\alpha C (X-1)}{X^{2}}=\frac{\alpha C
b x}{(1+bx)^2}
\end{equation}
where $\alpha$ is a constant has the advantage of  simplifying
(\ref{eq:b9}). The electric field given in (\ref{eq:b10}) vanishes
at the centre of the star, and remains continuous and bounded for
all interior points in the star. When $b=1$ then $E^2$ reduces to
the expression in the treatment of Maharaj and Komathiraj [17].
 Thus the choice for $E$ in (\ref{eq:b10}) is physically reasonable
in the study of the gravitational behaviour of charged stars. With
the help of (\ref{eq:b10}) we find that (\ref{eq:b9}) becomes
\begin{equation}
\label{eq:b11} 4X \left[a X-(a-b)\right]\frac{d^{2}Y}{dX^{2}}+
2(a-b)\frac{dY}{dX} + \left[ (b- a) -\alpha \right]Y = 0.
\end{equation}
 The differential equation (\ref{eq:b11}) becomes
\begin{equation}
\label{eq:b12} 4X \left[a X-(a-b)\right]\frac{d^{2}Y}{dX^{2}}+
2(a-b)\frac{dY}{dX} + (b- a)  Y = 0
\end{equation}
when $\alpha =0$ and there is no charge.

\section{Solutions} We need to integrate the master
equation (\ref{eq:b11}) to solve the Einstein-Maxwell system
(\ref{eq:b5a})-(\ref{eq:b5d}). Two categories of solution are
possible when $a=b$ and $a\neq b$.

\subsection{The Case $a=b$} When $a=b$  equation (\ref{eq:b11})
becomes
\begin{equation}
\label{eq:b13} X^2 \frac{d^{2}Y}{dX^{2}} - \frac{\alpha}{4a}Y = 0
\end{equation}
which is an Euler-Cauchy equation.  The solution of (\ref{eq:b13})
becomes
\begin{equation}
\label{eq:b14} Y = \left\{\begin{array}{ll} c_1 (
1+ax)^{(1+\sqrt{1+\alpha /a})/2}+
 c_2(1+ax)^{(1-\sqrt{1+\alpha /a})/2}  & \mbox{if $a> 0$}, \\
   & \\
 \sqrt{ 1+ax }\left[ c_1 \sin\left(\sqrt{ \frac{a+\alpha}{4a}}
  \ln (1+ax) \right) \right. \\
  \left.+c_2 \cos\left(\sqrt{ \frac{a+\alpha}{4a}}
  \ln (1+ax)\right)\right] & \mbox{if $a< 0$},
\end{array} \right.
\end{equation}
where $c_1$ and $c_2$ are constants. From (\ref{eq:b5a}) and
(\ref{eq:b6}) we observe that $\rho= - \frac{E^2}{2}$. We do not
pursue this case to avoid  negative energy densities. It is
interesting to observe that when $a=b=0$ then it is possible to
generate an exact Einstein-Maxwell solution to
(\ref{eq:b5a})-(\ref{eq:b5d}), for a different choice of $E^2$,
which contains the Einstein universe as pointed out by Komathiraj
and Maharaj [18].

\subsection{The Case $a \neq b$}
Observe that it is not
 possible to express the general solution of the master equation
 (\ref{eq:b11}) in terms of conventional
elementary functions for all values of $a, b~ (a\neq b)$ and $
\alpha$. In general the solution can be written  in terms of
special functions.  It is necessary to express  the solution in a
simple form so  that it is possible to conduct  a detailed
physical analysis.   Hence in this section we attempt to obtain a
general solution to the differential equation
 (\ref{eq:b11}) in series form. In a subsequent section we show
 that it is possible to find particular  solutions in terms
 of algebraic functions and polynomials.

We can utilise the method of Frobenius about $X=0$, since this is
a regular singular point of the differential equation
(\ref{eq:b11}). We write the solution of the differential equation
(\ref{eq:b11}) in the series  form
\begin{equation}
\label{eq:b15} Y = \sum_{n=0}^{\infty}c_{n}X^{n+r}, ~~ c_{0}\not=0
\end{equation}
where $c_{n}$ are the coefficients of the series and $r$ is a
constant. For an acceptable solution we need to find the
coefficients $c_{n}$ as well as the parameter ${r}$. On
substituting (\ref{eq:b15}) in the differential equation
(\ref{eq:b11}) we have
\begin{eqnarray}
2(a-b)c_{0}r[-2(r-1)+1]X^{r-1} +
\sum_{n=1}^{\infty} \left[2(a-b)c_{n+1}(n+r+1)[-2(n+r)+1] \right.&   & \nonumber \\
      &   & \nonumber \\
\label{eq:b16}
+c_{n}\left.[4a(n+r)(n+r-1)-(a-b+\alpha)]\right]X^{n+r}=0. & &
\end{eqnarray}
For consistency the coefficients of the various powers of $X$ must
vanish in (\ref{eq:b16}). Equating the coefficient of $X^{r-1}$ in
(\ref{eq:b16}) to zero, we find
\[ (a-b)c_{0}r[2(r-1)-1] = 0 \]
which is the indicial equation.  Since $c_{0}\not=0$ and $a\neq
b$, we must have $r=0$ or $r= \frac{3}{2}.$
 Equating the coefficient of $X^{n+r}$ in (\ref{eq:b16}) to zero we obtain
\begin{equation}
\label{eq:b17} c_{n+1}  =  \frac{4a(n+r)(n+r-1)- \left[a-b+\alpha
\right]}{2(a-b)(n+1+r)
[2(n+r)-1]}c_{n} , ~~ n\geq 0 \\
\end{equation}
The result (\ref{eq:b17}) is the basic difference equation which
determines the nature of the solution.

We can establish a general structure for all the coefficients by
considering the leading terms. We note that the coefficients $c_1,
c_2, c_3, ...$ can all be written in terms of the leading
coefficient $c_0$, and this leads to the expression
\begin{equation}
\label{eq:b18} c_{n+1} = \prod_{p=0}^{n}\frac{4
a(p+r)(p+r-1)-(a-b+\alpha)}{2(a-b)(p+1+r)[2(p+r)-1]}c_{0}
\end{equation}
where the symbol $\prod$ denotes multiplication. It is also
possible to establish  the result (\ref{eq:b18}) rigorously by
using the principle of mathematical induction. We can now generate
two linearly independent solutions
 from (\ref{eq:b15}) and (\ref{eq:b18}). For the parameter value $r=0$
 we obtain the first solution
\begin{eqnarray}
\label{eq:b19}Y_{1}& =& c_{0}
\left[1+\sum_{n=0}^{\infty}\prod_{p=0}^{n}\frac{4
a p(p-1) -(a-b+\alpha)}{2(a-b)(p+1)(2p-1)}X^{n+1} \right] \nonumber\\
& & \\
 y_{1} &=& c_{0} \left[1+\sum_{n=0}^{\infty}\prod_{p=0}^{n}\frac{4
a p(p-1) -(a-b+\alpha)}{2(a-b)(p+1)(2p-1)} (1+b x)^{n+1}
\right].\nonumber
\end{eqnarray}
For the parameter value $r=\frac{3}{2}$ we obtain the second
solution
\begin{eqnarray}
\label{eq:b20} Y_{2} &=& c_{0}X^{\frac{3}{2}}
\left[1+\sum_{n=0}^{\infty}
\prod_{p=0}^{n}\frac{a(2p+3)(2p+1)-(a-b+\alpha)}{(a-b)(2p+5)(2p+2)}X^{n+1}
\right] \nonumber \\
& & \\
 y_{2}& =& c_{0}(1+bx)^{\frac{3}{2}}
\left[1+\sum_{n=0}^{\infty}\prod_{p=0}^{n}\frac{a(2p+3)(2p+1)-
(a-b+\alpha)}{(a-b)(2p+5)(2p+2)}(1+bx)^{n+1} \right].\nonumber
\end{eqnarray}
Therefore the general solution to the differential equation
(\ref{eq:b7}),
 for the choice (\ref{eq:b10}), is given by
\begin{equation}
\label{eq:b21} y = a_1 y_{1}(x) + b_1y_{2}(x)
\end{equation}
where $a_1$ and $b_1$ are arbitrary constants and $y_{1}$ and
$y_{2}$ are given by (\ref{eq:b19}) and (\ref{eq:b20})
respectively.
 It is clear that the quantities $y_1$ and $y_2$ are linearly independent functions.
From (\ref{eq:b5a})-(\ref{eq:b5d}) and (\ref{eq:b21}) the general
solution to the Einstein-Maxwell system can be written as
\begin{subequations}
\begin{eqnarray}
\label{eq:b22a}
e^{2\lambda}   & = & \frac{1+bx}{1+a x} \\
\label{eq:b22b} e^{2\nu}       & = & A^{2}y^{2} \\
\label{eq:b22c} \frac{\rho}{C} & = & \frac{(b-a)(3+b x)}{(1+b
x)^{2}}
- \frac{\alpha b x}{2(1+b x)^{2}} \\
\label{eq:b22d} \frac{p}{C}    & = &
4\frac{(1+ax)}{(1+bx)}\frac{\dot{y}}{y}+
\frac{(a-b)}{(1+bx)} + \frac{\alpha bx}{2(1+bx)^{2}} \\
\label{eq:b22e} \frac{E^{2}}{C}& = & \frac{\alpha bx}{(1+bx)^{2}}.
\end{eqnarray}
\end{subequations}
The result in  (\ref{eq:b22a})-(\ref{eq:b22e}) is a new solution
to the Einstein-Maxwell field equations. Note that if we set
$\alpha=0$,
 (\ref{eq:b22a})-(\ref{eq:b22e}) reduce to models  for uncharged stars
 which may contain  new solutions to the Einstein field equations
  (\ref{eq:b4a})-(\ref{eq:b4c}).

\section{Elementary Functions}
 The general solution
(\ref{eq:b21}) can be expressed in terms of polynomial and
algebraic functions. This is possible in general because the
series (\ref{eq:b19}) and (\ref{eq:b20}) terminate for restricted
values of the parameters $a, b$ and $\alpha$ so that elementary
functions are possible. Consequently we obtain two sets of general
solutions in terms of elementary functions, by determining the
specific restriction on the quantity $a-b+\alpha$ for a
terminating series. The elementary functions found using this
method, can be  written as polynomials and polynomials with
algebraic functions. We provide the details of the process in the
Appendix; here we present a summary of the results. In terms of
the original variable $x$, the first category of solution can be
written as
\begin{eqnarray}
& & y  = \nonumber\\
&& d_1  (1+a x)^{\frac{1}{2}} \left[ 1 -(n+1)
 \sum_{i=1}^{n+1} \left(\frac{4a}{b-a}\right)^i \frac{(2i-1)(n+i)!}{(2i)!(n-i+1)!}
(1+bx)^{i}\right]    \nonumber \\
 \label{eq:b23}  &   & + d_2 \left(1+bx \right)^{\frac{3}{2}} \left[ 1
+ \frac{3}{(n+1)} \sum_{i=1}^{n}\left(\frac{4a}{b-a}\right)^i
\frac{(2i+2)(n+i+1)!}{(n-i)!(2i+3)!} (1+bx)^{i} \right]
\end{eqnarray}
for \(a-b +\alpha = a(2n+3)(2n+1)\), where $d_1$ and $d_2$ are
arbitrary constants. The second category of solutions can be
written as
\begin{eqnarray}
& & y  = \nonumber \\
& & d_3 (1+ax)^{\frac{1}{2}}
 (1+bx)^{\frac{3}{2}} \left[ 1+
\frac{3}{n(n-1)} \sum_{i=1}^{n-2}\left(\frac{4a}{b-a}\right)^i
\frac{(2i+2)(n+i)!}{(2i+3)!(n-i-2)!}
 (1+bx)^{i} \right] \nonumber \\
  &   & \nonumber \\
 \label{eq:b24} &   & + d_4 \left[1 - n(n-1)\sum_{i=1}^{n}\left(\frac{4a}{b-a}\right)^i
\frac{(2i-1)(n+i-2)!}{(2i)!(n-i)!} (1+bx)^{i} \right]
\end{eqnarray}
for  \(a-b+ \alpha = 4 a n(n-1)\), where $d_3$ and $d_4$ are
arbitrary constants. It is remarkable to observe that  the
solutions (\ref{eq:b23}) and (\ref{eq:b24}) are expressed
completely in terms of elementary functions only. This does not
happen often considering the nonlinearity of the gravitational
interaction in the presence of charge.  We have given our
solutions in a simple form: this has the advantage of facilitating
the analysis of the physical features of the stellar models.
Observe that our approach has combined both the charged and
uncharged cases for a relativistic star: when $\alpha=0$ we obtain
the solutions for the  uncharged case directly.

\section{Known Solutions} It is interesting to observe
that we can  regain a number of physically reasonable models  from
the general class of solutions found in this paper. These individual
models can be generated from the general series solution
(\ref{eq:b21}) or the simplified elementary functions (\ref{eq:b23})
and (\ref{eq:b24}). We explicitly generate the following models.

\subsection{Case I: Hansraj and Maharaj charged stars}
For this case we set  $a=0, b=1$ and $0 \leq \alpha < 1$. Then
from (\ref{eq:b19}) we find that
\begin{eqnarray}
 y_1 &= & c_0 \left[ 1 + \sum_{n=0}^{\infty}\prod_{p=0}^{n}
\frac{-(1-\alpha)}{2 (p+1)(2p-1)}(\sqrt{1+x})^{2n+2}\right] \nonumber\\
 &=&  c_0 \left(  \left[ 1 - \frac{(\sqrt{(1-\alpha)(1+x)})^2}{2!}
+ \frac{(\sqrt{(1-\alpha)(1+x)})^4}{4!}  \right. \right.\nonumber \\
 & & \left.- \frac{(\sqrt{(1-\alpha)(1+x)})^6}{6!} +...\right]
 + \sqrt{(1-\alpha)(1+x)}\left[\sqrt{(1-\alpha)(1+x)} \frac{}{} \right.
 \nonumber\\
&& \left. \left.-\frac{(\sqrt{(1-\alpha)(1+x)})^3}{3!}+
 \frac{(\sqrt{(1-\alpha)(1+x)})^5}{5!}-...
 \right]\right)\nonumber\\
 &=&  c_0 \cos{\sqrt{(1-\alpha)(1+x)}} + c_0
\sqrt{(1-\alpha)(1+x)}
 \sin{\sqrt{(1-\alpha)(1+x)}}. \nonumber
 \end{eqnarray}
Equation  (\ref{eq:b20}) gives the result
\begin{eqnarray}
y_2&= &c_0 (\sqrt{1+x})^3\left[ 1 +
\sum_{n=0}^{\infty}\prod_{p=0}^{n}
\frac{-(1-\alpha)}{(2p+5)(2p+2)}(\sqrt{1+x})^{2n+2}\right] \nonumber\\
&=& \frac{3c_0}{(\sqrt{1-\alpha})^3} \left( \left[
\sqrt{(1-\alpha)(1+x)}-\frac{(\sqrt{(1-\alpha)(1+x)})^3}{3!}
\right. \right. \nonumber\\
& & \left. + \frac{(\sqrt{(1-\alpha)(1+x)})^5}{5!}- ... \right]
 - \sqrt{(1-\alpha)(1+x)} \left[ 1 \frac{}{}  \right. \nonumber\\
 & &  \left.\left. - \frac{(\sqrt{(1-\alpha)(1+x)})^2}{2!}+
 \frac{(\sqrt{(1-\alpha)(1+x)})^4}{4!}
 - ...\right]\right) \nonumber\\
 &=& \frac{3 c_0}{(\sqrt{1-\alpha})^3}
\left[\sin\sqrt{(1-\alpha)(1+x)} -
\sqrt{(1-\alpha)(1+x)}\cos\sqrt{(1-\alpha)(1+x)}\right].\nonumber
\end{eqnarray}
Hence the general solution becomes
\begin{eqnarray}
 y &= &\left[D_1 - D_2
\sqrt{(1-\alpha)(1+x)}\right] \cos\sqrt{(1-\alpha)(1+x)} \nonumber\\
\label{eq:b25}& & + \left[D_2 + D_1
\sqrt{(1-\alpha)(1+x)}\right]\sin\sqrt{(1-\alpha)(1+x)}
\end{eqnarray}
where $D_1$ and $D_2$ are new arbitrary constants. The class of
charged solutions (\ref{eq:b25}) is the first category found by
Hansraj and Maharaj [19].

When $a=0, b=1$ and $ \alpha = 1$  we easily obtain the result
\begin{equation}
\label{eq:b26} y= a_1+b_1 (1+x)^{\frac{3}{2}}
\end{equation}
from (\ref{eq:b21}). This is the second category of the
Hansraj-Maharaj charged solutions.

We now set  $a=0, b=1$ and $\alpha > 1$. Then from (\ref{eq:b19})
we obtain
\begin{eqnarray}
y_1 &=& c_0 \left[ 1 + \sum_{n=0}^{\infty}\prod_{p=0}^{n}
\frac{(\alpha -1)}{2 (p+1)(2p-1)}(\sqrt{1+x})^{2n+2}\right] \nonumber\\
 &=& c_0 \left(  \left[ 1 + \frac{(\sqrt{(\alpha-1)(1+x)})^2}{2!}
  + \frac{(\sqrt{(\alpha -1)(1+x)})^4}{4!}
 \right. \right. \nonumber \\
 &  & \left. +\frac{(\sqrt{(\alpha -1)(1+x)})^6}{6!}+...\right]
 - \sqrt{(\alpha -1)(1+x)}\left[\sqrt{(\alpha -1)(1+x)} \right.\nonumber\\
 && \left. \left. +\frac{(\sqrt{(\alpha -1)(1+x)})^3}{3!}+
 \frac{(\sqrt{(\alpha -1)(1+x)})^5}{5!}+...
 \right]\right)\nonumber \\
 &=& c_0 \cosh{\sqrt{(\alpha -1)(1+x)}} - c_0 \sqrt{(\alpha -1)(1+x)}
  \sinh{\sqrt{(\alpha
 -1)(1+x)}}. \nonumber
 \end{eqnarray}
Equation  (\ref{eq:b20}) gives the result
\begin{eqnarray}
y_2 &= & c_0 (\sqrt{1+x})^3\left[ 1 +
\sum_{n=0}^{\infty}\prod_{p=0}^{n}
\frac{(\alpha -1)}{(2p+5)(2p+2)}(\sqrt{1+x})^{2n+2}\right] \nonumber\\
&=& \frac{- 3c_0}{(\sqrt{\alpha -1})^3} \left( \left[
\sqrt{(\alpha -1)(1+x)}+ \frac{(\sqrt{(\alpha -1)(1+x)})^3}{3!}\right. \right.\nonumber\\
 & & \left. + \frac{(\sqrt{(\alpha -1)(1+x)})^5}{5!}+ ... \right]-
\sqrt{(\alpha -1)(1+x)} \left[ 1 \right. \nonumber\\
& & \left. \left.+ \frac{(\sqrt{(\alpha -1)(1+x)})^2}{2!} +
\frac{(\sqrt{(\alpha -1)(1+x)})^4}{4!} + ...\right]\right) \nonumber\\
 &=& \frac{- 3c_0}{(\sqrt{\alpha
-1})^3}(\sinh\sqrt{(1-\alpha)(1+x)} -
\sqrt{(1-\alpha)(1+x)}\cosh\sqrt{(1-\alpha)(1+x)}). \nonumber
\end{eqnarray}
Therefore,  the general solution becomes
\begin{eqnarray}
 y &= &\left[D_2 - D_1
\sqrt{(\alpha -1)(1+x)}\right] \sinh\sqrt{(\alpha -1)(1+x)} \nonumber\\
\label{eq:b27}& & + \left[D_1-D_2 \sqrt{(\alpha
-1)(1+x)}\right]\cosh\sqrt{(\alpha -1)(1+x)}
\end{eqnarray}
where $D_1$ and $D_2$ are new arbitrary constants. This is the third
category of charged solutions found by Hansraj and Maharaj.

The exact solutions (\ref{eq:b25}), (\ref{eq:b26}) and
(\ref{eq:b27})  were comprehensively studied by Hansraj and Maharaj
[19], and it was shown that these solutions correspond to a charged
relativistic sphere which is realistic as all conditions for
physically acceptability are met. The condition of  causality is
satisfied and the speed of light is greater than the speed of sound.

\subsection{Case II: Maharaj and Komathiraj charged stars}
If $b=1$, then (\ref{eq:b23}) becomes
\begin{eqnarray}
y & = & d_1  (1+a x)^{\frac{1}{2}} \left[ 1 -(n+1)
 \sum_{i=1}^{n+1} \left(\frac{4a}{1-a}\right)^i \frac{(2i-1)(n+i)!}{(2i)!(n-i+1)!}
(1+x)^{i}\right]    \nonumber \\
 \label{eq:b28}  &   & + d_2 \left(1+x \right)^{\frac{3}{2}} \left[ 1
+ \frac{3}{(n+1)} \sum_{i=1}^{n}\left(\frac{4a}{1-a}\right)^i
\frac{(2i+2)(n+i+1)!}{(n-i)!(2i+3)!} (1+x)^{i} \right]
\end{eqnarray}
for $a-1+\alpha =a(2n+1)(2n+3)$. When $b=1$ then  (\ref{eq:b24})
gives
\begin{eqnarray}
y & = & d_3 (1+a x)^{\frac{1}{2}}
 (1+x)^{\frac{3}{2}} \left[ 1+
\frac{3}{n(n-1)} \sum_{i=1}^{n-2}\left(\frac{4a}{1-a}\right)^i
\frac{(2i+2)(n+i)!}{(2i+3)!(n-i-2)!}
 (1+x)^{i} \right] \nonumber \\
  &   & \nonumber \\
 \label{eq:b29} &   & + d_4 \left[1 - n(n-1)\sum_{i=1}^{n}\left(\frac{4a}{1-a}\right)^i
\frac{(2i-1)(n+i-2)!}{(2i)!(n-i)!} (1+x)^{i} \right]
\end{eqnarray}
for $a-1+\alpha=4an(n-1)$. The two categories of  solutions
(\ref{eq:b28}) and (\ref{eq:b29}) correspond to the Maharaj and
Komathiraj [17] model for a compact sphere in electric fields. The
Maharaj and Komathiraj charged stars have a simple form in terms of
elementary functions; they are physically reasonable and contain the
Durgapal and Bannerji [16] model and other exact models
corresponding to neutron stars.

\subsection{Case III: Finch and Skea neutron stars}
When $\alpha=0$, we obtain
\begin{equation}
\label{eq:b30} y=\left[D_1 - D_2 \sqrt{1+x}\right]\cos\sqrt{1+x}+
\left[D_2 + D_1 \sqrt{1+x}\right]\sin\sqrt{1+x}
\end{equation}
from (\ref{eq:b25}). Thus, we regain the  Finch and Skea [20] model
for a neutron star when the electromagnetic field is absent. The
Finch and Skea neutron star model has been shown to satisfy all the
physical criteria for an isolated spherically symmetric stellar
uncharged source. It is for this reason that this model has been
used by many researchers to model the interior of neutron stars.

\subsection{Case IV: Durgapal and Bannerji neutron stars}
If we take $\alpha =0$ and  $n=0$ then  $2a+b =0$, and  we get
\[ y= d_1 (1+ax)^{\frac{1}{2}} (5 -4ax)+ d_2
(1-2ax)^{\frac{3}{2}}\] from (\ref{eq:b23}). If we set $a=-
\frac{1}{2}$ $(ie.~ b=1)$, then it is easy to verify that this
equation becomes
\begin{equation}
\label{eq:b31} y =c_{1}(2-x)^{\frac{1}{2}}(5+2x)+
c_{2}(1+x)^{\frac{3}{2}}
\end{equation}
where $c_{1}= d_1/3\sqrt{2}$ and $c_{2}=d_2$ are new arbitrary
constants. Thus we have regained the Durgapal and Bannerji [16]
neutron star model. This model satisfies all physical criteria for
acceptability and has been utilised by many researchers to model
uncharged neutron stars.

\subsection{Case V: Tikekar Superdense Stars}
If we take $\alpha =0$ and $n=2$ then  $7a+ b=0$, and we find
\[ y = d_3(1+ax)(1-7ax)^{\frac{3}{2}} + d_4 \left[ 1+ \frac{1}{2}
(1-7ax)-\frac{1}{8} (1-7ax)^2\right] \] from (\ref{eq:b24}). If we
set $a=-1$ $(ie.~b=7)$ and let $\tilde{x}=\sqrt{1-x}$ then this
equation  becomes
\begin{equation}
\label{eq:b32} y= c_1 \tilde{x}
(1-\frac{7}{8}\tilde{x}^2)^{\frac{3}{2}} + c_2 \left[1 -
\frac{7}{2}\tilde{x}^2 + \frac{49}{24} \tilde{x}^4 \right]
\end{equation}
where $c_1= d_3 8^{\frac{3}{2}}$ and $c_2 = - d_4 /3$ are new
arbitrary constants. Thus we have regained the Tikekar [21] model
for superdense neutron star from our general solution. The Tikekar
superdense model plays an important role in describing highly
dense matter, cold compact matter and core-envelope models for
relativistic stars. The Tikekar relativistic star falls into a
more general class of models with spheroidal spatial geometry
found by Maharaj and Leach [12]; this class can be generalised to
include the presence of an electric field as shown by Komathiraj
and Maharaj [22].

\section{Physical Analysis}
In this section we demonstrate that the exact solutions found in
this paper are physically reasonable and may be used to model a
charged relativistic sphere.  We observe from (\ref{eq:b22a}) and
(\ref{eq:b22b}) that the gravitational potentials $e^{2\nu}$ and
$e^{2\lambda}$ are continuous in the stellar interior and nonzero
at the centre for all values of the parameters $a, b$ and
$\alpha$. From (\ref{eq:b22c}), we can express the variable  $x$
in terms of the energy density $\rho$ only as
\[x= \frac{1}{4 b} \left\{ C[2(b-a) -\alpha]\rho^{-1}
 \pm \sqrt{C^2[2(b-a)-\alpha]^2 \rho^{-2}+8C[4(b-a)+\alpha]\rho^{-1}}-4 \right\}.\]
Therefore from (\ref{eq:b22d}), the isotropic pressure $p$ can  be
express in terms of $\rho$ only. Thus all the forms of the solutions
presented in this paper satisfy the barotropic equation of state
$p=p(\rho)$ which is a desirable feature. Note that many of the
solutions appeared in the literature do not satisfy this property.

To illustrate the graphical behaviour of the matter variables in
the stellar interior we consider the particular solution
(\ref{eq:b25}). In this case the line element becomes
\begin{eqnarray}
ds^2 &=& -A^2 \left[\left(D_1 -D_2
\sqrt{(1-\alpha)(1+r^2)}\right)\cos\sqrt{(1-\alpha)(1+r^2)}\right.
\nonumber\\
& & + \left. \left(D_2 + D_1
\sqrt{(1-\alpha)(1+r^2)}\right)\sin\sqrt{(1-\alpha)(1+r^2)}\right]^2
dt^2 \nonumber\\
\label{eq:b33} & & +(1+ Cr^2)dr^2 +r^2(d\theta^2 +\sin^2\theta
d\phi^2).
\end{eqnarray}
For simplicity we make the choice $A=1, C=1, D_1=1, D_2= 4$ in the
metric (\ref{eq:b33}). We choose $\alpha=\frac{1}{2}$ for the
charged and we consider the  interval $0\leq r \leq 1$ to generate
the relevant plots.

We utilised the software package Mathematica to generate the plots
for $\rho, p, E^2 $ and $\frac{dp}{d\rho}$ respectively. The dotted
line corresponds to $\alpha =\frac{1}{2}$ and $E^2\neq 0$; the solid
line corresponds to $\alpha =0$ and $E^2 =0$. In Fig.1 we have
plotted the energy density on the interval $0\leq r \leq 1$. It can
be easily seen that the energy densities in both cases are positive
and continuous at the centre; it is a monotonically decreasing
function throughout the interior of the star from centre to the
boundary. In Fig.2 we have plotted the behaviour of the isotropic
pressure. The pressure $p$ remains regular in the interior and is
monotonically decreasing. The role of the electromagnetic field is
highlighted in figures 1 and 2: the effect of $E^2$ is to produce
smaller values for $\rho$ and $p$. From figures 1 and 2 we observe
that the presence of $E$ does not significantly affect $\rho$ but
has a much greater influence on $p$ closer to the centre. We believe
that this follows directly from our choice (\ref{eq:b10}) for the
electric field intensity; other choices of $E$ generate different
profiles as indicated in [18].
 The electric field intensity $E^2$ is given in Fig.3 which is
positive, continuous and monotonically increasing. In Fig.4 we
have plotted $\frac{dp}{d\rho}$ on the interval $0\leq r \leq 1$
for both charged and uncharged cases. We  observe that
$\frac{dp}{d\rho}$ is always positive and less than unity. This
indicates that the speed of the sound is less than the speed of
the light and causality is maintained. Note that the effect of the
electromagnetic field is to produce lower values for
$\frac{dp}{d\rho}$ and the speed of sound is decreased when
$\alpha \neq 0$. Hence we have shown that the solution
(\ref{eq:b25}), for our particular chosen parameter values,
satisfies the requirements for a physically reasonable charged
body.

\vspace{0.5cm}
\begin{figure}[thb]
\vspace{1.5in} \includegraphics{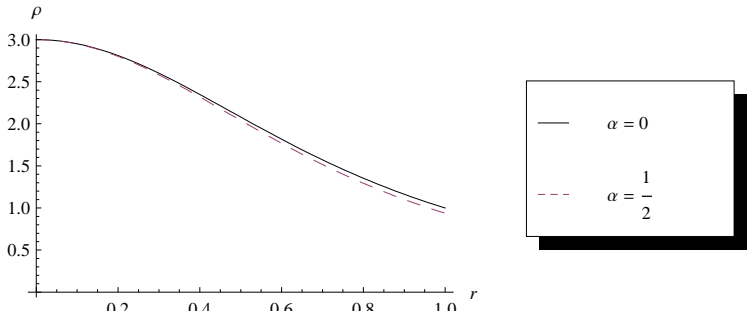}
\caption{\label{Graph-FS-Roa} Energy density in geometrised units.}
\end{figure}

\vspace{1cm}

\begin{figure}[thb]
\vspace{1.5in} \includegraphics{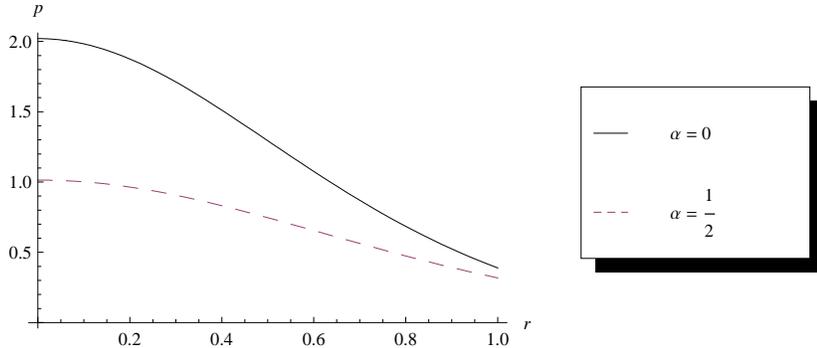}
\caption{\label{Graph-FS-p-3} Matter pressure in geometrised units.}
\end{figure}

\vspace{1.5cm}

\begin{figure}[thb]
\vspace{2.5in} \includegraphics{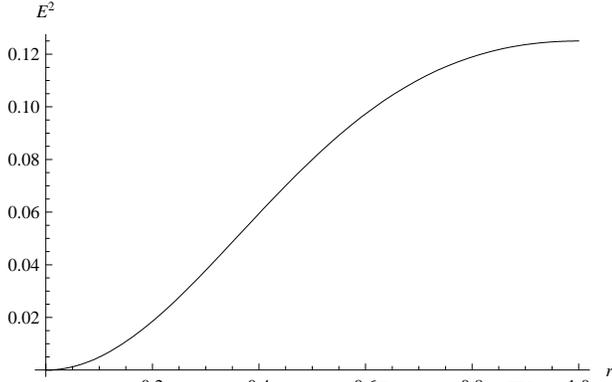}
\caption{\label{Graph-FS-E}Electric field in geometrised units.}
\end{figure}

\vspace{1.5cm}

\begin{figure}[thb]
\vspace{1.5in} \includegraphics{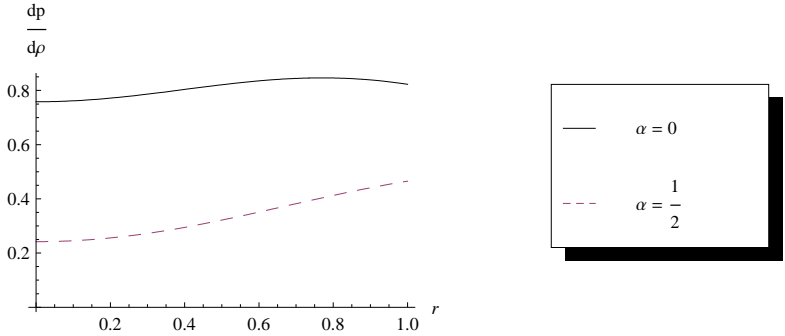}
\caption{\label{Graph-FS-Roa-p-2} The gradient $dp/d\rho$ in
geometrised units.}
\end{figure}

\section{Discussion}
In this paper we have found a new class of exact solutions
(\ref{eq:b21}) to the Einstein-Maxwell system of equations
(\ref{eq:b5a})-(\ref{eq:b5d}). We demonstrated that  solutions
(\ref{eq:b23}) and (\ref{eq:b24}) in terms of elementary functions
can be extracted from the general class. If $\alpha=0$ then we
regain uncharged solutions of the Einstein field equations which may
be new. The charged solutions of  Hansraj and Maharaj [19] and
Maharaj and Komathiraj [17] were shown to be special cases of our
general class; the uncharged neutron star models of Finch and Skea
[20], Durgapal and Bannerji [16] and Tikekar [21] are regained when
the electromagnetic field vanishes. The simple form of the exact
solutions in terms of polynomials and algebraic functions
facilitates the analysis of the physical features of a charged
sphere. The solutions found satisfy a barotropic equation of state.
For particular parameter values we plotted the behaviour of $\rho,
p, E^2 $ and $\frac{dp}{d\rho}$ and showed that they were physically
reasonable. A pleasing feature of our plots is that we can
distinguish between charged and uncharged exact solutions. The
presence of charge leads to smaller values of $\rho, p $ and
$\frac{dp}{d\rho}$ in the figures generated. This indicates that the
presence of charge can dramatically affect the behaviour of the
matter and gravitational variables.

\section*{Acknowledgements}
ST thanks the National Research Foundation and the University of
KwaZulu-Natal for financial support, and is grateful to Eastern
University, Sri Lanka for study leave. SDM acknowledges that this
work is based upon research supported by the South African
Research Chair Initiative of the Department of Science and
Technology and the National Research Foundation.\\

\section*{Appendix}
In the Appendix we derive the elementary functions (\ref{eq:b23})
and (\ref{eq:b24}).

On substituting $r=0$ into equation (\ref{eq:b17}) we find that
\begin{equation}
\label{eq:b38} c_{i+1}  =
\frac{4ai(i-1)-((a-b)+\alpha)}{(a-b)(2i+2)(2i-1)}c_{i} ,
 ~~ i\geq 0.
\end{equation}
If we set $a-b +\alpha =4an(n-1)$,  where $n$ is a fixed integer
and we assume that $a\neq 0$, then $c_{n+1}=0$.  It is easy to see
that the subsequent coefficients $c_{n+2}, c_{n+3}, c_{n+4},...$
vanish and equation (\ref{eq:b38}) has the solution
\begin{equation}
\label{eq:b39} c_i= - n(n-1)\left(\frac{4a}{b-a}\right)^i
\frac{(2i-1)(n+i-2)!} {(2i)!(n-i)!}c_{0} ~ , 1\leq i \leq n.
\end{equation}
Then from equation (\ref{eq:b15}) (when $r=0$) and (\ref{eq:b39})
we generate
\begin{equation}
\label{eq:b40} Y_{1} = c_{0} \left[1- n(n-1)\sum_{i=1}^{n} \left(
\frac{4a}{b-a}\right)^i \frac{(2i-1)(n+i-2)!}{(2i)!(n-i)!}X^{i}
\right]
\end{equation}
where \( a-b+\alpha=4a n(n-1). \)

On substituting $r = \frac{3}{2}$ into (\ref{eq:b17}), we get
\begin{equation}
\label{eq:b41} c_{i+1}  =  \frac{a(2i+3)(2i+1)-((a-b)+\alpha)}
{(a-b)(2i+5)(2i+2)}c_{i} , ~~ i\geq 0.
\end{equation}
If we set $a-b+\alpha =a(2n+3)(2n+1)$, where $n$ is a fixed
integer and we assume that $a\neq 0$, then $c_{n+1}=0$. Also we
see that the subsequent coefficients  $c_{n+2}, c_{n+3}, c_{n+4},
...$ vanish and equation (\ref{eq:b41}) is solved  to give
\begin{equation}
\label{eq:b42}
 c_i =\left( \frac{4a}{b-a}\right)^i \frac{3 (2i+2)(n+i+1)!
}{(n+1)(n-i)!(2i+3)!}c_{0},~~~~1 \leq i \leq n.
\end{equation}
Then from  equations (\ref{eq:b15}) (when $ r=\frac{3}{2}$ ) and
(\ref{eq:b42}) we generate
\begin{equation}
\label{eq:b43} Y_{1} = c_{0}X^{\frac{3}{2}} \left[ 1 +
\frac{3}{(n+1)} \sum_{i=1}^{n} \left(\frac{4a}{b-a} \right)^i
\frac{(2i+2)(n+i+1)!}{(n-i)!(2i+3)!}X^{i} \right]
\end{equation}
where $a-b+\alpha = a(2n+3)(2n+1)$.
 The  elementary functions (\ref{eq:b40}) and (\ref{eq:b43})
comprise the first solution of the differential equation
(\ref{eq:b11}) for appropriate values of $a-b+\alpha$.

We take the second solution of (\ref{eq:b11}) to be of the form
\[Y_2 = \left[a X -(a-b)\right]^{\frac{1}{2}} u(X)\]
where $u(X)$ is an arbitrary function.  On substituting $Y_2$ into
(\ref{eq:b11}) we obtain
\begin{equation}
\label{eq:b44} 4 X \left[a X -(a-b)\right] \ddot{u} - 2 \left[2 aX
+ (a-b)\right]\dot{u} - \left[2a -b +\alpha\right]u=0
\end{equation}
where dots denote differentiation with respect to $X$. We write
the solution of the differential equation (\ref{eq:b44}) in the
series form
\begin{equation}
\label{eq:b45} u = \sum_{n=0}^{\infty}c_{n}X^{n+r} ~,~~~~~~c_{0}
\not= 0.
\end{equation}
On substituting (\ref{eq:b45}) into the differential equation
(\ref{eq:b44})
 we find
\begin{eqnarray}
2(a-b)c_{0}r[-2(r-1)+1]X^{r-1} -  \sum_{n=0}^{\infty}
\left(2(a-b)c_{n+1}(n+1+r)[2(n+r)-1] \right. & & \nonumber \\
 \left. - c_{n}[4 a(n+r)^{2} - (2a
-b+\alpha)]\right) X^{n+r} & = & 0. \nonumber\\
\label{eq:b46} & &
\end{eqnarray}
Setting the coefficient of $X^{r-1}$ in (\ref{eq:b46}) to zero we
find
 \[ (a-b)c_{0}r[2(r-1)-1] = 0. \]
which is the indicial equation. Since $c_{0}\not= 0$ and $a\neq b$
we must have $r=0$ or $r= \frac{3}{2}$. Equating the coefficient
of $X^{n+r}$ in (\ref{eq:b46}) to zero we find that
\begin{equation}
\label{eq:b47} c_{n+1}  =  \frac{4a(n+r)^{2}
-(2a-b+\alpha)}{2(a-b)(n+r+1)[2(n+r)-1]} c_{n}
\end{equation}
We establish a general structure for the coefficients by considering
the leading terms.

On substituting $r=0$ in equation (\ref{eq:b47}) we obtain
\begin{equation}
\label{eq:b48} c_{i+1}  =  \frac{4 ai^{2} -(2a
-b+\alpha)}{(a-b)(2i+2)(2i-1)} c_{i}.
\end{equation}
We assume that \(a-b + \alpha = a(2n+3)(2n+1)\) where $n$ is a
fixed integer. Then $c_{n+2}=0$ from (\ref{eq:b48}). Consequently
the remaining coefficients  $c_{n+3}, c_{n+4}, c_{n+5}, ... $
vanish and equation (\ref{eq:b48}) has the solution
\begin{equation}
\label{eq:b49}
 c_i= -(n+1)\left(\frac{4a}{b-a}\right)^i \frac{(2i-1)(n+i)!}{(2i)!(n-i+1)!}c_{0} ,~~1 \leq i \leq n+1.
\end{equation}
Then from the equations (\ref{eq:b45}) (when $r = 0$) and
(\ref{eq:b49}) we find
\[ u = c_{0} \left[ 1-   (n+1)
\sum_{i=1}^{n+1}\left(\frac{4a}{b-a}\right)^i
\frac{(2i-1)(n+i)!}{(2i)!(n-i+1)!}X^{i} \right]. \] Hence we
generate the result
\begin{equation}
\label{eq:b50} Y_{2} = c_{0}\left[aX -(a-b)\right]^{\frac{1}{2}}
\left[ 1- (n+1) \sum_{i=1}^{n+1}\left(\frac{4a}{b-a}\right)^i
\frac{(2i-1)(n+i)!}{(2i)!(n-i+1)!}X^{i}\right]
\end{equation}
where $a-b + \alpha =a(2n+3)(2n+1).$

On substituting \( r= \frac{3}{2}\) into equation (\ref{eq:b47})
we obtain
\begin{equation}
\label{eq:b51} c_{i+1}= \frac{a(2i+3)^{2}-(2a
-b+\alpha)}{(a-b)(2i+5)(2i+2)}c_{i}.
\end{equation}
We assume that \( a-b+\alpha = 4an(n-1)\) where $n$ is a fixed
integer. Then $c_{n-1}=0$ from (\ref{eq:b51}).   Consequently the
remaining coefficients  $c_{n}, c_{n+1}, c_{n+2}, ...$ vanish and
(\ref{eq:b51}) can be solved to obtain
\begin{equation}
\label{eq:b52} c_i= \left(\frac{4a}{b-a}\right)^i
\frac{3(2i+2)(n+i)!}{n(n-1)(2i+3)!(n-i-2)!}c_{0},~~i \leq n-2.
\end{equation}
Then from the equations (\ref{eq:b45}) (when $r= \frac{3}{2}$) and
(\ref{eq:b52}) we have
\[u = c_{0} X^{\frac{3}{2}} \left[ 1+ \frac{3}{n(n-1)} \sum_{i=1}^{n-2}\left(\frac{4a}{b-a}\right)^i
\frac{(2i+2)(n+i)!}{(2i+3)!(n-i-2)!}X^{i} \right]. \] Hence we
generate the result
\begin{equation}
\label{eq:b53} Y_{2} = c_{0}\left[aX -(a-b)\right]^{\frac{1}{2}}
X^{\frac{3}{2}} \left[ 1+ \frac{3}{n(n-1)}
\sum_{i=1}^{n-2}\left(\frac{4a}{b-a}\right)^i
\frac{(2i+2)(n+i)!}{(2i+3)!(n-i-2)!}X^{i} \right]
\end{equation}
where $a-b + \alpha = 4 a n(n-1).$ The functions (\ref{eq:b50}) and
(\ref{eq:b53}) generate the second solution of the differential
equation (\ref{eq:b11}).

The solutions found can be written in terms of two classes of
elementary functions. We have the first category of solutions
\begin{eqnarray}
Y & = & D_1 \left[a X -(a-b)\right]^{\frac{1}{2}} \left[ 1-(n+1)
\sum_{i=1}^{n+1}\left(\frac{4a}{b-a}\right)^i
\frac{(2i-1)(n+i)!}{(2i)!(n-i+1)!}X^{i}\right]    \nonumber \\
  &   & \nonumber \\
  \label{eq:b54}&   & + D_2 X^{\frac{3}{2}} \left[ 1 + \frac{3}{(n+1)}
   \sum_{i=1}^{n}\left(\frac{4a}{b-a}\right)^i
\frac{(2i+2)(n+i+1)!}{(n-i)!(2i+3)!}X^{i} \right]
\end{eqnarray}
for \(a-b +\alpha = a(2n+3)(2n+1)\), where $D_1$ and $D_2$ are
arbitrary constants. In terms of $x$ the solution (\ref{eq:b54})
becomes
\begin{eqnarray}
& y =& d_1  (1+a x)^{\frac{1}{2}} \left[ 1 -(n+1)
 \sum_{i=1}^{n+1} \left(\frac{4a}{b-a}\right)^i \frac{(2i-1)(n+i)!}{(2i)!(n-i+1)!}
(1+bx)^{i}\right]    \nonumber \\
 \label{eq:b55}  & +  d_2& \left(1+bx \right)^{\frac{3}{2}} \left[ 1
+ \frac{3}{(n+1)} \sum_{i=1}^{n}\left(\frac{4a}{b-a}\right)^i
\frac{(2i+2)(n+i+1)!}{(n-i)!(2i+3)!} (1+bx)^{i} \right]
\end{eqnarray}
where $d_1 =D_1\sqrt{b}$ and $d_2=D_2$ are new arbitrary
constants. The second category of solutions is given by
\begin{eqnarray}
Y & = & D_1  \left[ a X-(a-b))\right]^{\frac{1}{2}}X^{\frac{3}{2}}
\left[ 1+ \frac{3}{n(n-1)}
 \sum_{i=1}^{n-2} \left(\frac{4a}{b-a}\right)^i \frac{(2i+2)(n+i)!}{(2i+3)!(n-i-2)!}X^{i} \right] \nonumber \\
  &   & \nonumber \\
\label{eq:b56}&   & + D_2 \left[1-
n(n-1)\sum_{i=1}^{n}\left(\frac{4a}{b-a}\right)^i
\frac{(2i-1)(n+i-2)!}{(2i)!(n-i)!}X^{i} \right]
\end{eqnarray}
for  \(a-b+ \alpha = 4 a n(n-1)\), where $D_1$ and $D_2$ are
arbitrary constants. In terms of $x$ the solution (\ref{eq:b56})
becomes
\begin{eqnarray}
y & = & d_3 (1+ax)^{\frac{1}{2}}
 (1+bx)^{\frac{3}{2}} \left[ 1+
\frac{3}{n(n-1)} \sum_{i=1}^{n-2}\left(\frac{4a}{b-a}\right)^i
\frac{(2i+2)(n+i)!}{(2i+3)!(n-i-2)!}
 (1+bx)^{i} \right] \nonumber \\
  &   & \nonumber \\
 \label{eq:b57} &   & + d_4 \left[1 - n(n-1)\sum_{i=1}^{n}\left(\frac{4a}{b-a}\right)^i
\frac{(2i-1)(n+i-2)!}{(2i)!(n-i)!} (1+bx)^{i} \right]
\end{eqnarray}
where $d_3 =D_1\sqrt{b}$ and $d_4=D_2$ are new arbitrary
constants.\\

~\\
{\Large{\bf References}}
\begin{enumerate}
\item Ivanov BV. Static charged perfect fluid sphere in general
relativity. \emph{Physical Review D} 2002; \textbf{65}:104001.

\item Sharma R, Mukherjee S, Maharaj SD. General solution for a
class of static charged spheres. \emph{General Relativity and
Gravitation} 2001; \textbf{33}:999-1009.

\item Patel LK, Koppar SS. A charged analogue of the
Vaidya-Tikekar solution. \emph{Australian Journal of Physics}
1987; \textbf{40}:441-447.

\item Tikekar R, Singh GP. Interior Reissner-Nordstrom metric on
spheroidal space-times. \emph{Gravitation and Cosmology} 1998;
\textbf{4}:294-296.

\item Mukherjee B. Static spherically symmetric perfect fluid
distribution in presence of electromagnetic field. \emph{Acta
Physica  Hungarica} 2001; \textbf{13}:243-252.

\item Gupta YK, Kumar M. A superdense star model as charged
analogue of Schwarzschild's interior solution.  \emph{General
Relativity and Gravitation} 2005; \textbf{37}:575-583.

\item Thomas VO, Ratanpal BS, Vinodkumar P C. Core-envelope models
of superdense star with anisotropic envelope. \emph{International
Journal of modern Physics D} 2005; \textbf{14}:85-96.

\item Tikekar R, Thomas VO. Relativistic fluid sphere on
pseudo-spheroidal space-time.  \emph{Pramana Journal of Physics}
1998; \textbf{50}:95-103.

\item Paul BC, Tikekar R. A core-envelope model of compact stars.
 \emph{Gravitation and Cosmology} 2005; \textbf{11}:244-248.

\item Mak MK, Harko T. Quark stars admitting a one-parameter group
of conformal motion.  \emph{International Journal of Modern
Physics D} 2004; \textbf{13}: 149-156.

\item Komathiraj K, Maharaj SD. Analytical models for quark stars.
\emph{International Journal of  Modern Physics D} 2007;
\textbf{16}:1803-1811.

\item Maharaj SD, Leach PGL. Exact solutions for the Tikekar
superdense star. \emph{Journal of Mathematical Physics} 1996;
\textbf{37}:430-437.

\item Thirukkanesh S, Maharaj SD. Exact models for isotropic
matter. \emph{Classical and Quantum Gravity} 2006;
\textbf{23}:2697-2709.

\item Maharaj SD, Thirukkanesh S. Generating potentials via
difference equations. \emph{Mathematical Methods in the Applied
Sciences} 2006; \textbf{29}:1943-1952.

\item John AJ, Maharaj SD. An exact isotropic solution.  \emph{Il
Nuovo Cimento B} 2006; \textbf{121}:27-33.

\item Durgapal MC, Bannerji R. New analytical stellar model in
general relativity. \emph{Physical Review D} 1983;
\textbf{27}:328-331.

\item Maharaj SD, Komathiraj K. Generalized compact spheres in
electric field.  \emph{Classical and Quantum Gravity} 2007;
\textbf{24}:4513-4524.

\item Komathiraj K, Maharaj SD. Classes of exact Einstein-Maxwell
solutions. \emph{General Relativity and Gravitation} 2007;
\textbf{39}:2079-2093.

\item Hansraj S, Maharaj SD. Charged analogue of Finch-Skea stars.
 \emph{International Journal of  Modern Physics D} 2006; \textbf{15}:1311-1327.

\item Finch MR, Skea JEF. A realistic stellar model based on an
ansatz of Duorah and Ray.  \emph{Classical and Quantum Gravity}
1989; \textbf{6}:467-476.

\item Tikekar R. Exact model for a relativistic star.
\emph{Journal of Mathematical Physics} 1990;
\textbf{31}:2454-2458.

\item Komathiraj K Maharaj SD. Tikekar superdense stars in
electric fields. \emph{Journal of  Mathematical Physics} 2007;
\textbf{48}:042501.

\end{enumerate}

\end{document}